# Hybrid Electro-Optic Modulator Combining Silicon Photonic Slot Waveguides with High-k Radio-Frequency Slotlines

Sandeep Ummethala[1,2,6], Juned N. Kemal[1], Ahmed S. Alam[1], Matthias Lauermann[1], Yasar Kutuvantavida[1], Sree H. Nandam[3], Lothar Hahn[2], Delwin L. Elder[5], Larry R. Dalton[5], Thomas Zwick[4], Sebastian Randel[1], Wolfgang Freude[1], Christian Koos[1,2,7]

[1]*Karlsruhe Institute of Technology (KIT), Institute of Photonics and Quantum Electronics (IPQ), 76131 Karlsruhe, Germany*
[2]*Karlsruhe Institute of Technology (KIT), Institute of Microstructure Technology (IMT), 76344 Eggenstein-Leopoldshafen, Germany*
[3]*Karlsruhe Institute of Technology (KIT), Institute of Nanotechnology (INT), 76344 Eggenstein-Leopoldshafen, Germany*
[4]*Karlsruhe Institute of Technology (KIT), Institute of Radio Frequency Engineering and Electronics (IHE), 76131 Karlsruhe, Germany*
[5]*University of Washington, Department of Chemistry, Seattle, Washington 98195, USA*
[6]*sandeep.ummethala@kit.edu,* [7]*christian.koos@kit.edu*

**Abstract:** Electro-optic (EO) modulators rely on interaction of optical and electrical signals with second-order nonlinear media. For the optical signal, this interaction can be strongly enhanced by using dielectric slot-waveguide structures that exploit a field discontinuity at the interface between a high-index waveguide core and the low-index EO cladding. In contrast to this, the electrical signal is usually applied through conductive regions in the direct vicinity of the optical waveguide. To avoid excessive optical loss, the conductivity of these regions is maintained at a moderate level, thus leading to inherent *RC*-limitations of the modulation bandwidth. In this paper, we show that these limitations can be overcome by extending the slot-waveguide concept to the modulating radio-frequency (RF) signal. Our device combines an RF slotline that relies on $BaTiO_3$ as a high-k dielectric material with a conventional silicon photonic slot waveguide and a highly efficient organic EO cladding material. In a proof-of-concept experiment, we demonstrate a 1 mm-long Mach-Zehnder modulator that offers a 3 dB-bandwidth of 76 GHz and a 6 dB-bandwidth of 110 GHz along with a small π-voltage of 1.3 V ($U_\pi L$ = 1.3 V mm). To the best of our knowledge, this represents the largest EO bandwidth so far achieved with a silicon photonic modulator based on dielectric waveguides. We further demonstrate the viability of the device in a data transmission experiment using four-state pulse-amplitude modulation (PAM4) at line rates up to 200 Gbit/s. Our first-generation devices leave vast room for further improvement and may open an attractive route towards highly efficient silicon photonic modulators that combine sub-1 mm device lengths with sub-1 V drive voltages and modulation bandwidths of more than 100 GHz.

## 1. Introduction

High-speed electro-optic (EO) modulators are key devices for optical communications [1–3], optical metrology [4], microwave photonics [5], or ultra-broadband signal processing at THz bandwidths [6]. Ideal modulators should combine small π-voltages $U_\pi$ and short device lengths $L$ with large modulation bandwidths, while offering a path to cost-efficient mass production and monolithic co-integration with advanced photonic circuitry. In practice, however, it is challenging to fulfill all these requirements simultaneously. When it comes to scalability and high-density integration, the silicon photonic (SiP) platform would be the technology of choice, exploiting highly mature CMOS processes and offering a rich portfolio of advanced photonic devices that can be realized with high yield on large-area wafers [7]. However, due to the absence of the Pockels effect in silicon, conventional SiP modulators rely on carrier injection or depletion in *p-n* junctions that are integrated into the optical waveguides [8]. This leads to an inherent trade-off between device efficiency and modulation bandwidth. As an example, depletion-type SiP modulators were demonstrated with bandwidths of 48 GHz, but the efficiency of these devices was rather low with $U_\pi L$ products of 7.4 V mm [9]. The efficiency of SiP modulators can be greatly improved by combining low-loss slot waveguides on silicon-on-insulator (SOI) with optimized organic electro-optic materials in a hybrid approach [10]. These so-called silicon-organic hybrid (SOH) devices can benefit from ultra-high in-device electro-optic coefficients of, e.g., 390 pm/V, leading to ultra-low voltage-length products down to $U_\pi L$ = 0.32 V mm [11]. However, without a supporting gate voltage [12], the EO bandwidth of highly efficient slot-waveguide SOH modulators is limited [13] to, e.g., 25 GHz or less [14] due to the *RC* time constant associated with the capacitance of the slot and the resistance of the adjacent doped Si slabs [13]. Plasmonic-organic hybrid (POH) MZM [6,15–17] can overcome these limitations by replacing the silicon slabs with highly conductive gold pads and by exploiting surface plasmon polaritons in the resulting metallic slot waveguide. This, however, comes at the price of substantial optical losses $a$, leading to rather high loss-efficiency products [18] $aU_\pi L$ of more than 10 V dB [6,19] as compared to 1 V dB for SOH devices [11,20]. Other integration platforms such as thin-film lithium-niobate or indium phosphide (InP) have shown impressive modulation bandwidths of up to 100 GHz [21], but their efficiency is limited by rather high voltage-length-products $U_\pi L$ in excess of 6 V mm [22], and monolithic co-integration with other devices is difficult. Modulators based on indium-phosphide (InP) can also offer large bandwidths of, e.g., 80 GHz [22], but the voltage length-products of more than 6 V mm are still comparatively high, and fabrication relies on rather expensive processes that cannot compete with the scalability and maturity of the silicon photonic platform. Thus, a scalable approach to realize highly efficient and low-loss modulators with large electro-optic bandwidth is still lacking.

In this paper, we report on a novel concept for hybrid silicon photonic modulators that allows combining high efficiency of organic electro-optic materials with large modulation bandwidths without the need for lossy plasmonic structures. The device overcomes the *RC* limitations of conventional SOH slot-waveguide modulators by replacing resistive coupling through the doped Si slabs with a capacitive coupling via high-k dielectric material [23]. In a proof-of-concept experiment, we demonstrate a 1 mm-long capacitively-coupled SOH (CC-SOH) Mach-Zehnder modulator that exploits barium titanate ($BaTiO_3$) as high-k dielectric and simultaneously offers a large EO bandwidth of 76 GHz and a small π-voltage of 1.3 V ($U_\pi L = 1.3\,\mathrm{V\,mm}$). To the best of our knowledge, this represents the largest EO bandwidth achieved with a silicon photonic modulator based on dielectric waveguides. To validate the viability of the device, we generate a 4-state pulse-amplitude modulation (PAM4) signals at symbol rates (line rates) up to 100 GBd (200 Gbit/s) with bit-error ratios (BER) below the threshold for soft-decision forward-error correction. As our first-generation devices are not yet optimized and leave vast room for further improvement, we believe that the CC-SOH concept opens an attractive route towards highly efficient silicon photonic modulators that combine sub-1 mm device lengths with sub-1 V drive voltages, sub-1 dB phase-shifter losses, and modulation bandwidths of 100 GHz or more.

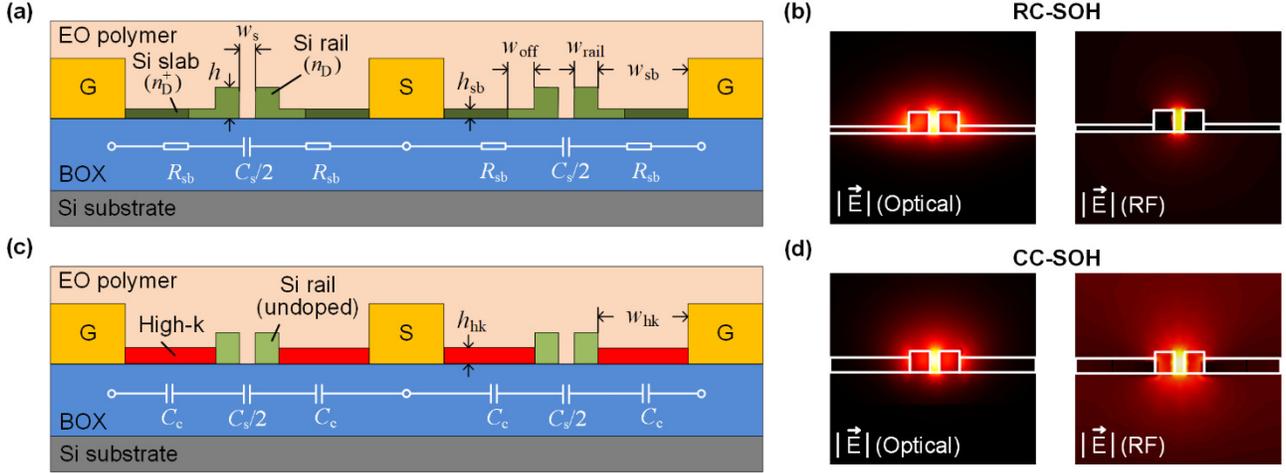

**Fig. 1.** Resistively-coupled vs. capacitively-coupled silicon-organic hybrid (SOH) modulator. **(a)** Cross-section of a conventional resistively-coupled SOH (RC-SOH) Mach-Zehnder modulator (MZM) on the silicon-on-insulator (SOI) platform. The phase modulator in each arm of the MZM comprises a silicon slot waveguide formed by two silicon (Si) rails (green) with widths $w_{\mathrm{rail}} = 150\ldots200\,\mathrm{nm}$, which are separated by a gap of width $w_{\mathrm{s}} = 80\ldots120\,\mathrm{nm}$. The device is covered by an organic EO material (light orange), which also fills the slot. The Si rails are connected to the metal strips of the ground-signal-ground (GSG) transmission line (yellow) through doped Si slabs (green) such that, at low frequencies, the RF modulating voltage drops predominantly in the slot region. To reduce optical loss, a smaller doping concentration $n_{\mathrm{D}}$ is used in the rails and in the directly adjacent slab regions of width $w_{\mathrm{off}}$ (light green), whereas the slab regions further away from the optical slot waveguide (dark green) are subjected to a higher doping concentration $n_{\mathrm{D}}^+$. A simplified equivalent-circuit model of the RC-SOH MZM is depicted as a white overlay. In this model, the slot waveguide of each phase modulator is represented by a capacitance $C_{\mathrm{s}}/2$, which is connected to the ground and the signal strip of the transmission line through a total slab resistance $2R_{\mathrm{sb}}$. **(b)** Electric-field profiles of the optical quasi-TE mode and the RF mode of the RC-SOH phase modulator. Both figures indicate the magnitudes of the complex electric field vectors. The optical and the RF field show strong field overlap, resulting in a high modulation efficiency. **(c)** Cross-section of capacitively-coupled SOH (CC-SOH) MZM where the doped Si slabs in (a) are substituted by a high-k dielectric material (red), thus replacing the resistive coupling by a capacitive coupling. As depicted in the simplified equivalent circuit, the slot capacitance $C_{\mathrm{s}}/2$ of each MZM arm is electrically connected to the transmission line electrodes through a coupling capacitance $C_{\mathrm{c}}$. The coupling capacitance $C_{\mathrm{c}}$ is typically much larger than the slot capacitance $C_{\mathrm{c}} \gg C_{\mathrm{s}}$, such that RF voltage applied to the transmission line drops predominantly across the slot. In contrast to RC-SOH modulators, CC-SOH devices do not suffer from optical loss in the Si rails. **(d)** Electric-field profiles of the optical quasi-TE mode and the RF mode of the CC-SOH phase modulator. Since the relative permittivity of the high-k dielectric is much larger than the relative permittivity of the Si or the EO polymer, the RF field in the slot region is strongly enhanced. The structure thus combines a silicon photonic slot waveguide for optical frequencies and a high-k dielectric slotline for RF frequencies. The strong confinement of both the optical and RF fields to the EO material in the slot region ensures efficient modulation.

## 2. SOH Modulator: Resistive vs Capacitive coupling

The concept of capacitively-coupled silicon-organic hybrid (CC-SOH) modulators and their advantages over resistively-coupled SOH (RC-SOH) devices is explained in Fig. 1. Figure 1(a) shows the cross section of a conventional RC-SOH MZM [18], realized on a silicon-on-insulator (SOI) substrate. Each arm of the MZM comprises a silicon (Si) slot waveguide formed by two Si rails (green) with typical widths $w_{\mathrm{rail}} = 150\ldots200\,\mathrm{nm}$, which are separated by a slot of width $w_{\mathrm{s}} = 80\ldots120\,\mathrm{nm}$. The radio-frequency (RF) modulating signal is carried by coplanar transmission line electrodes (yellow) in ground-signal-ground (GSG) configuration. The Si rails are connected to the metal transmission line via doped Si slabs (dark green) having a height $h_{\mathrm{sb}} = 50\ldots70\,\mathrm{nm}$ and a width $w_{\mathrm{sb}} = 1.2\ldots2.0\,\mathrm{\mu m}$. The slot region is filled with an organic EO material which has a refractive index ($n_{\mathrm{EO}} \approx 1.75$), much smaller than that of Si ($n_{\mathrm{Si}} \approx 3.5$). This leads to a pronounced field enhancement of the optical quasi-transverse-electric (quasi-TE) mode in the slot region [24]. At sufficiently low modulation frequencies, the voltage applied to the transmission line entirely drops across the narrow slot region. This leads to a tight confinement of the electric RF field to the slot and ensures strong overlap with the optical mode, which results in high modulation efficiency, see Fig. 1(b). The concept of RC-SOH MZM opens a path towards compact devices with sub-1mm phase shifter lengths and sub-1V operation voltages [11,20] that can be directly driven by highly scalable CMOS circuits without the need for a separate amplifier [25]. At high operation frequencies, however, the slot capacitance cannot be fully charged and discharged through the resistive slabs during one modulation cycle [13]. For a quantitative description of the RC-SOH MZM dynamics, we use a simple equivalent-circuit model, illustrated as a white overlay in Fig. 1(a). Each of the two slot waveguides is represented by a slot capacitance $C_s/2$, which is connected to the metal transmission line through a total slab resistance $2R_{sb}$. Taking into account the contribution of both arms, this results in an intrinsic $RC$-limited bandwidth $f_{RC} = 1/(2\pi R_{sb} C_s)$ for the RC-SOH MZM. High-bandwidth RC-SOH devices require low slab resistivity and hence high doping concentration, which leads to larger optical losses. This trade-off between optical loss and $RC$-limited bandwidth can be avoided by using a tailored doping profile, where the Si rails and the directly adjacent slab regions have a lower doping concentrations $n_D$ while much higher concentrations $n_D^+$ are used further away from the slot [13], see Fig. 1(a). Still, the $RC$-limited bandwidth turns out to be one of the most stringent restrictions of experimentally demonstrated RC-SOH devices [14,26,27]. In laboratory experiments, this limitation could only be overcome by applying a relatively high gate voltage $U_{gate}$ across the buried $SiO_2$ (BOX) layer [12] to induce a charge accumulation layer in the Si slabs. However, this gate voltage typically exceeds 100 V [12,14] and is thus not a solution for practical devices.

CC-SOH MZM overcomes these limitations by avoiding the resistive slabs altogether and by using capacitive coupling instead. To this end, the doped Si slabs in Fig. 1(a) are replaced by a high-k dielectric material, which forms a large coupling capacitor between the silicon rails and the metal strips of the GSG transmission line, see Fig. 1(c). The simplified equivalent circuit of this scheme is depicted as a white overlay in Fig. 1(c). Each phase modulator of the MZM is modeled by a slot capacitance $C_s/2$, which is connected to the metal transmission line through a coupling capacitance $C_c$ to each side of the slot. If the high-k dielectric has a relative permittivity $\varepsilon_r$ much larger than that of Si and the EO polymer such that $C_c \gg C_s$, then the RF electric field drops predominantly across the slot region. The high-k dielectric is chosen such that the refractive index for optical wavelengths is smaller than $n_{Si}$ such that confinement of the optical mode to the slot region is not impaired. Figure 1(d) shows the electric-field profiles of the optical quasi-TE mode and RF mode of the CC-SOH phase modulator. Since the relative permittivity of the high-k dielectric is much larger than the relative permittivity of the Si or the EO polymer, the electric RF field in the slot region is strongly enhanced. The structure thus combines a silicon photonic slot waveguide for optical frequencies with a high-k dielectric slotline for RF frequencies [28]. The strong confinement of both the optical and RF fields to the EO material in the slot region ensures efficient modulation, without the need for any doping of the Si rails. By proper choice of high-k materials with low absorption in the near infrared [28], CC-SOH devices should hence permit simultaneous realization of low loss and large modulation bandwidths. Examples of high-k dielectric materials with $\varepsilon_r > 100$ and refractive indices smaller than $n_{Si}$ are $TiO_2$, $SrTiO_3$, $BaSrTiO_3$ and $BaTiO_3$. Since the capacitive coupling between the transmission-line electrodes and the slot of a CC-SOH MZM is not associated with a time constant, the bandwidth is only limited by the frequency-dependent propagation loss of the modulating RF signal, impedance mismatch and the velocity mismatch between the RF and the optical waves.

## 3. Demonstration of CC-SOH MZM bandwidth

To demonstrate the viability of the CC-SOH concept, we fabricate a 1 mm-long CC-SOH MZM with $BaTiO_3$ (BTO) as high-k dielectric. The device is realized on an SOI substrate having a 220 nm-thick Si device layer and a 2 µm-thick buried oxide ($SiO_2$), see Supplementary Information 1, Section 1 for details. A schematic layout of the MZM is depicted in Fig. 2(a). Light is coupled to the SiP chip via an on-chip grating coupler (GC). A multimode interference (MMI) coupler splits the incoming light and launches it into the two arms of an unbalanced MZM. The two arms have a path difference of 80 µm that allow for an adjustment of the operating point by tuning the wavelength. The modulating RF signal is coupled to the MZM through a coplanar transmission line, which is realized in a ground-signal-ground (GSG) configuration. A second MMI at the other end of the MZM combines the modulated light and feeds it to an output waveguide, which is connected to another GC. Figure 2(b) shows a false-colored scanning electron microscope (SEM) image of a section of the CC-SOH MZM as defined by the dashed (green) rectangle in the Fig. 2(a). Each MZM arm consists of a Si strip waveguide (width $w = 500\,\text{nm}$; height $h = 220\,\text{nm}$), which is transformed into a slot waveguide using a strip-to-slot converter [29]. The slot waveguide comprises two Si rails (green) with widths $w_{rail} = 200\,\text{nm}$, which are separated by a slot of width $w_s = 100\,\text{nm}$, see inset in Fig. 2(b). A coplanar transmission line made from 150 nm-thick gold electrodes (yellow) in ground-signal-ground (GSG) configuration carries the RF signal. Amorphous BTO slabs (red) having a height $h_{hk} = 150\,\text{nm}$ and a width $w_{hk} = 1\,\mu\text{m}$ are deposited between the Si slot waveguide and the metal electrodes using room-temperature RF magnetron sputtering. We measure a refractive index $n_{BTO} = 1.85$ at a wavelength of 1550 nm for the amorphous BTO thin film, see Supplementary Information 1, Section 1A for details. After fabrication of the device, an organic EO polymer (YLD124) is filled into the silicon slots and subsequently activated by a one-time poling process [18]. To this end, the chip is heated above the glass transition temperature of the EO polymer, and a DC voltage is applied across the floating ground electrodes to align the dipolar chromophores in the two slots. The orientation of the chromophores is frozen by cooling the chip to room temperature while maintaining the DC poling field. This leads to a configuration where the driving electric RF field applied to the GSG transmission line is parallel to the poling direction in one MZM arm and antiparallel in the other arm, thereby enabling operation of CC-SOH MZM in the push-pull mode. In order to quantify the modulation efficiency, we measure the $\pi$-voltage $U_\pi$ of the MZM by driving it with a low-frequency triangular signal $U_d$ as depicted in Fig. 2(c), green curve. The intensity-modulated output (red curve) of the MZM is detected using a photodiode and is recorded by an oscilloscope along with the drive signal. We measure the voltage difference $U_\pi = 1.3\,\text{V}$ that is needed to drive the transmission of the MZM from its minimum to its maximum by introducing phase shift of $\pi$ between the optical signals in the 1 mm-long MZM arms. From the $U_\pi L$ product of 1.3 V mm, we estimate an electro-optic coefficient of $r_{33} \approx 34\,\text{pm/V}$ (assuming $n_{EO,YLD124} = 1.75$), see Supplementary Information 1, Section 2A for details.

To measure the frequency response of the CC-SOH MZM, the wavelength is adjusted for operation in the quadrature point. An RF drive signal with frequency in the range between 0.01 GHz and 110 GHz is supplied by a vector network analyzer (VNA, Keysight PNA-X N5247) and is coupled to the GSG transmission line of the MZM using a microwave probe. A 50 Ω impedance terminates the other end of the transmission line.

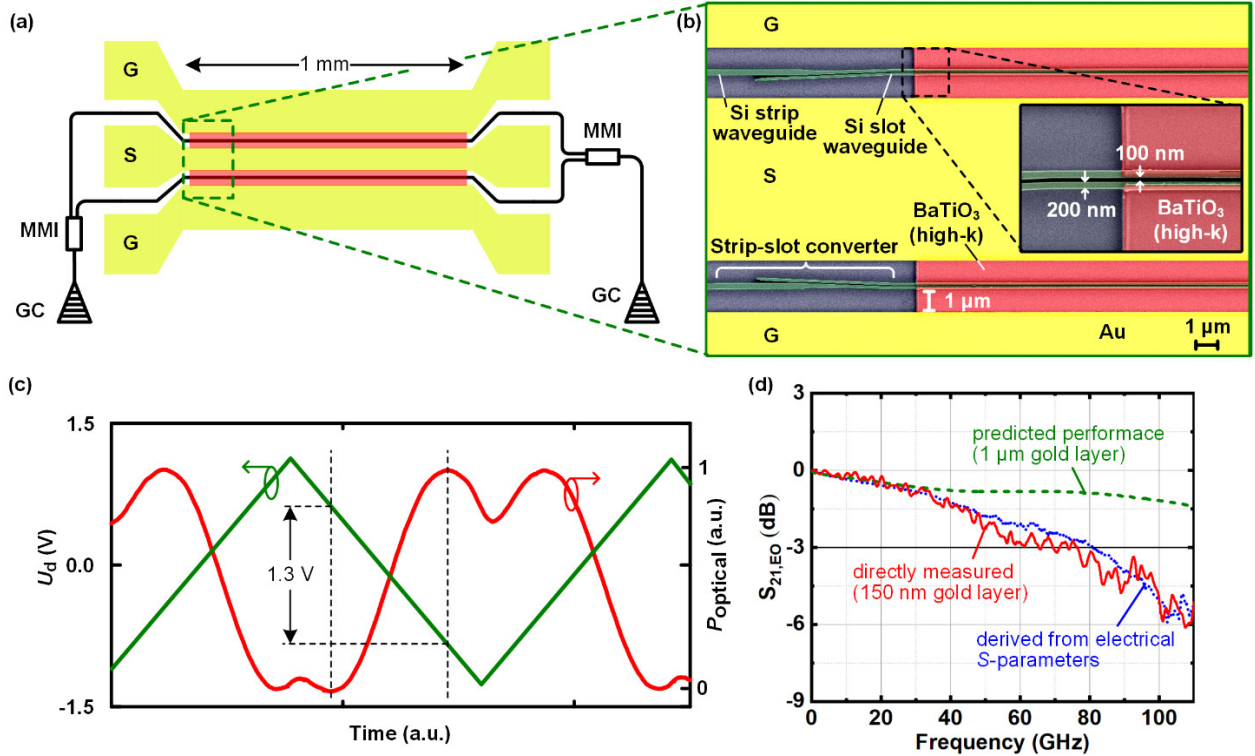

**Fig. 2**. Capacitively-coupled SOH MZM using barium titanate (BaTiO$_3$, BTO) as a high-k dielectric. **(a)** Schematic of a 1-mm-long CC-SOH MZM driven by a coplanar ground-signal-ground (GSG) transmission line. Light is coupled in and out of the device via on-chip grating couplers (GC). Multimode interference (MMI) couplers are used to split and recombine the light of the two MZM arms. **(b)** False-colored SEM image of a CC-SOH MZM section, as defined by the dashed green rectangle in (a). Each phase-shifter arm of the MZM comprises a strip-slot converter (green) for coupling light from the Si strip waveguide to the Si slot waveguide. A coplanar ground-signal-ground (GSG) transmission line made from gold (yellow) carries the RF signal. Amorphous BTO slabs (red) having a height of 150 nm and a width of 1 µm are deposited between the Si slot waveguide and the gold strips of the coplanar transmission line. The phase shifter section of each arm is 1 mm long. The inset shows a zoom-in of one of the MZM arms comprising a Si slot waveguide with a 100 nm-wide slot and a Si rail having a width of 200 nm. **(c)** Measurement of the π-voltage of a BTO-based CC-SOH MZM showing the intensity-modulation at the device output (red) for a low-frequency triangular drive signal (green). The π-voltage $U_\pi = 1.3\,\text{V}$ can be directly read from the voltage increment needed to drive the intensity modulation from minimum to maximum transmission. **(d)** Dynamic behavior of a 1-mm-long CC-SOH MZM with YLD124 as EO cladding material: The red curve shows the electro-optic (EO) response $S_{21,\text{EO}}$ measured using a vector network analyzer (VNA) and a calibrated high-speed photodiode. The measured 3 dB EO bandwidth of the 1 mm-long CC-SOH MZM is 76 GHz. As a reference, we also derive the EO response from the measured electrical $S$-parameters of the device using an analytical model [30], see Supplementary Information 1, Section 2B for details. The analytically derived result is indicated by the blue curve and agrees well with the directly measured behavior. The strong frequency-dependent decay of the EO response is mainly caused by RF propagation loss of the modulating signal along the coplanar transmission line having a thickness of only 150 nm. By using thicker gold strips, the bandwidth of the MZM can be significantly increased. This is indicated by the green curve, which corresponds to the predicted EO response when using 1 µm-thick gold strips for the coplanar transmission line. In this case, the expected 3 dB EO bandwidth extends far beyond 100 GHz, see Supplementary Information 1, Section 2B for details.

The intensity-modulated output of the CC-SOH MZM is detected with a calibrated high-speed photodiode (HHI, C05-W36) having a 3 dB bandwidth of 78 GHz, and its output signal is recorded by the VNA. The measured data is corrected by accounting for the frequency response of the photodiode, the probes, and the RF cables, leading to the electro-optic (EO) response $S_{21,\text{EO}}$ of the CC-SOH MZM, see red curve in Fig. 2(d). For a 1-mm-long BTO-based CC-SOH MZM, we measure a 3 dB EO bandwidth of approximately 76 GHz. Note that the 3 dB EO bandwidth corresponds to a frequency at which the phase-modulation index is reduced by a factor of $1/\sqrt{2}$, corresponding to a 3 dB decay in the associated RF power generated by the photodiode. Alternatively, the bandwidth can be specified in terms of the 6 dB bandwidth, which corresponds to a power decay of the associated photocurrent by a factor of four and amounts to approximately 110 GHz for the current device. In the literature [13,31–35] of EO modulators, both conventions of bandwidth are used, often without an explicit mention of the adopted specification.

Note that the bandwidth measured for the current MZM is not limited by the device concept but is a consequence of the non-optimum electrical design of our first-generation structures. Especially, the metal transmission line is fabricated using a lift-off process with a thin photoresist, which limited the thickness of the gold layer to a rather small value of approximately 150 nm. This leads to significant RF propagation loss, which increases strongly with frequency. To better understand the underlying bandwidth limitations, we perform a detailed evaluation [36] of the device dynamics. In first step, we measure the electrical $S$-parameters of the CC-SOH modulator, in particular the frequency-dependent complex amplitude reflection factor $\underline{S}_{11}$ at the input and the complex amplitude transmission factor $\underline{S}_{21}$. From the measured $S$-parameters, we then derive the line impedance, the RF propagation loss, and the RF propagation constant, which is associated with the RF effective refractive index $n_{e,RF}$, see Supplementary Information 1, Section 2B for details. Adopting the analytical model described in [30], we predict the EO response of the CC-SOH device which would be expected due to the electrical behavior, see Supplementary Information 1, Section 2B for details. This analysis includes the contribution due to the losses of the underlying RF transmission line, impedance mismatch, as well as the walk-off between the RF modulation signal and the optical signal. The results of this analysis, indicated as a blue trace in Fig. 2(d), coincide well with the directly measured EO response of the device, red trace, and we conclude that the decay of the EO response must have its origin in the electrical behavior of the RF line. We also find that walk-off between the RF signal and the optical signal does not represent a significant limitation of the current devices – based on an optical group refractive index $n_{g,opt} = 2.8$ at 1550 nm and an RF effective index $n_{e,RF} = 2.2$, we estimate a walk-off-related 3 dB bandwidth of approximately 220 GHz [37], see Supplementary Information 1, Section 2B for details. As a consequence, high RF propagation loss in combination with a frequency-dependent line impedance remains as the most important reason for the strong decay in the frequency response of our first-generation CC-SOH devices.

The RF propagation loss has two main contributions – conductor loss and dielectric loss. In order to investigate the contribution from the dielectric loss of BTO, we analyze the RF properties of BTO thin films that are deposited by room-temperature magnetron sputtering. We measure a relative permittivity $\varepsilon_{r,BTO} = 18$ and a loss tangent $\tan\delta_{BTO} = 0.05$ at 60 GHz, which is in good agreement with measurements of amorphous BTO films [38–41]. Details of the RF characterization of BTO thin films can be found in Supplementary Information 1, Section 1B. At 50 GHz, this leads to a BTO-related dielectric loss of 0.58 dB/mm, which is much smaller than the overall RF propagation loss of 4.5 dB/mm measured at the same frequency. We hence conclude that the ohmic loss of the 150 nm-thick transmission line dominates the RF propagation losses, which is also confirmed by an electrical simulation (CST MicrowaveStudio) of the 150-nm-thick gold transmission line assuming a conductivity of $1.65 \times 10^7 \, \text{S/m}$, see Fig. S8 in Supplementary Information 1, Section 2B. For thicker gold layers, e.g., 1 µm, the RF propagation loss can be significantly reduced to about 1 dB/mm at 50 GHz, see Fig. S8 in Supplementary Information 1, Section 2B. This would then allow extending the 3 dB bandwidth of the 1 mm-long CC-SOH MZM far beyond 100 GHz, see green curve in Fig. 2(d).

Besides increased modulation bandwidth, our first-generation CC-SOH modulators also leave room for improving the modulation efficiency and for reducing the optical loss. For our current devices, we measure an in-device EO coefficient of $r_{33} \approx 34$ pm/V. This value is far below the 230 pm/V demonstrated [42] for conventional RC-SOH MZM using similar EO material (YLD124) [43]. This is because the device is poled with a rather weak electric field. The poling field was limited by the voltage supplied by our current poling setup, which is designed for conventional RC-SOH devices where the effective electrode spacing corresponds to the gap width of typically 100 … 200 nm. In addition, we may use more efficient EO materials such as JRD1 [44], for which in-device EO coefficients of 390 pm/V have been demonstrated in conventional RC-SOH devices with 160 nm-wide slots [11]. We would hence expect that the in-device EO coefficient of CC-SOH devices could be improved by approximately an order of magnitude, thereby enabling $U_\pi L$ products well below 1 V mm, i.e., sub-1V drive voltages for sub-1 mm phase shifter lengths. Similarly, the phase shifters of our first-generation CC-SOH devices exhibit rather high propagation losses of approximately 6 dB/mm due to non-optimized fabrication processes, which lead to substantial sidewall roughness of the slot waveguides. With improved fabrication processes based on 193 nm deep-UV lithography, it should be possible to reduce these values to less than 0.2 dB/mm [45]. In combination with low-loss strip-to-slot converters (0.02 dB loss) [29] and MMI couplers (0.2 dB loss) [46], we would therefore expect on-chip insertion losses of less than 1 dB for fully optimized CC-SOH modulators.

## 4. High-speed signaling Experiments

Figure 3(a) shows the setup for evaluating the performance of a CC-SOH MZM in high-speed data transmission. The electrical drive signals are synthesized by a 120 GSa/s arbitrary-waveform generator (AWG, Keysight M8194A) with an analog bandwidth of 45 GHz. We use a pseudo-random bit sequence (length $2^{11} - 1$) along with pulse shapes featuring a raised-cosine power spectrum to generate two- and four-level electrical signals at symbol rates of 64 GBd and 100 GBd. The drive signals are fed to an RF amplifier (Centellax, UA1L65VM) before being coupled to the GSG pads of the CC-SOH MZM via a microwave probe (Cascade Microtech, i67A, 67 GHz bandwidth). The other end of the GSG transmission line is terminated with a 50 Ω impedance via a second microwave probe. In addition to the wavelength adjustment of the operating point in the unbalanced CC-SOH MZM, a DC voltage $U_{bias}$ is applied via a bias tee for fine-tuning. Digital pre-compensation is used to counteract the strong frequency roll-off of the AWG and other RF components used in the setup, excluding the CC-SOH MZM itself. The optical carrier at a wavelength near 1550 nm is provided by an external-cavity laser (ECL) and is adjusted for polarization using a fiber-based polarization controller (PC). Light is coupled in and out of the CC-SOH MZM via on-chip grating couplers (GC). The intensity-modulated output of the MZM is amplified by an erbium-doped fiber amplifier (EDFA), and the out-of-band amplified spontaneous emission (ASE) noise is suppressed by an optical band-pass filter (Opt. BPF) with a 2 nm passband. The signal is finally fed to a high-speed photodiode having a bandwidth of 70 GHz (Finisar XPDV3120R). The resulting electrical signal is boosted using an RF amplifier and is recorded by a 100 GHz real-time oscilloscope (Keysight UXR10004A). Figure 3(b) shows the eye diagrams and the histograms of the detected amplitudes at the sampling point for symbol rates of 64 GBd and 100 GBd. For 64 GBd on-off keying (OOK), no bit errors could be measured in our 15 µs-long recordings, which contain approximately $10^6$ symbols. We hence estimate a bit error ratio (BER) below $10^{-6}$. At 100 GBd OOK (100 Gbit/s line rate), a BER of $4 \times 10^{-6}$ is measured which is well below the hard-decision forward error correction [47] (HD-FEC) limit with 7 % overhead. For 64 GBd four-state pulse amplitude modulation (PAM4), we measure a BER of $5 \times 10^{-5}$, which is also below the FEC hard-decision FEC threshold with 7 % overhead. For 100 GBd PAM4 signaling with a line rate of 200 Gbit/s, we obtain a BER of $9 \times 10^{-3}$, which is below the threshold for soft-decision FEC (SD-FEC) limit with 20 % overhead. At 100 GBd, the peak-to-peak on-drive was estimated to be about 1 V. We find that the increase in BER for 100 GBd PAM4 signaling is mainly caused by the low quality of the electrical drive signal, for which a BER of $1.7 \times 10^{-3}$ was measured in an electrical back-to-back experiment, in which to output of the RF amplifier was connected to the real-time oscilloscope via an RF attenuator.

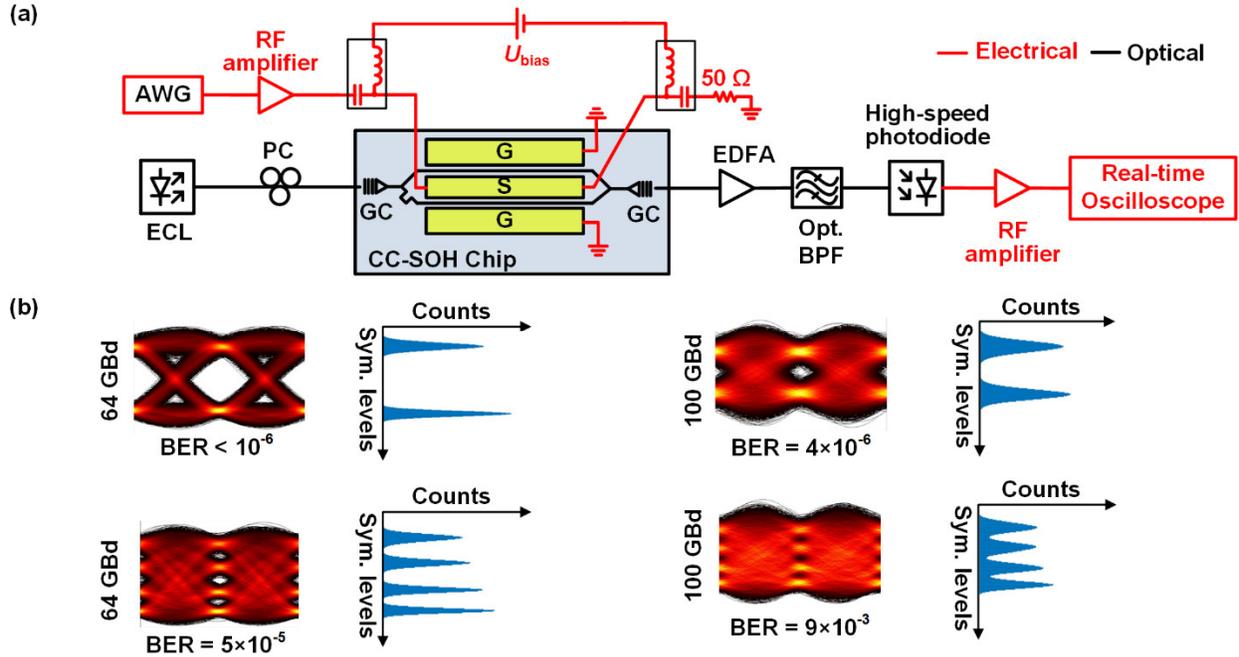

**Fig. 3.** High-speed signaling with a CC-SOH MZM using BTO as a high-k dielectric. **(a)** Setup for data signal generation and detection. An arbitrary-waveform generator (AWG) followed by an RF amplifier are used to drive the modulator. The electrical drive signal is fed to the ground-signal-ground (GSG) coplanar transmission line via a microwave probe (not shown). The operating point of the MZM with un-balanced arm lengths is set by the optical wavelength and fine-tuned by a DC voltage $U_{bias}$ applied through a bias tee. Light from an external-cavity laser (ECL) is coupled to the CC-SOH MZM via a polarization controller (PC) and an on-chip grating coupler (GC). An erbium-doped fiber amplifier (EDFA) is used to amplify the light from the output GC, and a band-pass filter (Opt. BPF) suppresses the amplified spontaneous emission (ASE) noise, before a high-speed photodiode receives the light. The RF signal from the photodiode is amplified and captured by a real-time oscilloscope. **(b)** Eye diagrams for symbol rates of 64 GBd and 100 GBd along with the histograms of the amplitudes in the sampling point and with measured bit error ratios (BER). At 64 GBd, the BER is below $10^{-6}$ for on-off keying (OOK) and reaches $5 \times 10^{-5}$ for four-state pulse amplitude modulation (PAM4). At 100 GBd OOK, a BER of $4 \times 10^{-6}$ is measured, well below the forward error correction (FEC) limit for 7 % overhead. For 100 GBd PAM4 (line rate 200 Gbit/s), we measure a BER of $9 \times 10^{-3}$, which is still below the threshold for soft-decision FEC (SD-FEC) limit with 20 % overhead.

## 5. Summary

We demonstrated a novel concept for SOH electro-optic modulators that relies on a capacitive coupling scheme to overcome the intrinsic bandwidth limitation of conventional devices with resistively-coupled slot waveguides. In a proof-of-concept experiment using $BaTiO_3$ as a high-k dielectric for enhanced capacitive coupling, we demonstrated a CC-SOH MZM having a 3 dB EO bandwidth of 76 GHz. To the best our knowledge, this is the largest bandwidth so far achieved for a Si photonic modulator based on dielectric waveguides. The device features a small π-voltage-length-product $U_\pi L = 1.3\,\mathrm{Vmm}$. The viability of the CC-SOH modulator is demonstrated in a high-speed data transmission experiment, in which we generate PAM4 signals at line rates up to 200 Gbit/s. These results obtained with first-generation devices leave a vast room for further improvements. We believe that the CC-SOH concept offers an attractive route towards highly efficient silicon photonic modulators that combine sub-1 mm device lengths with sub-1 V drive voltages, sub-1 dB insertion losses, and modulation bandwidths of 100 GHz or more.

**Funding**. This work is supported by the European Research Council (ERC) consolidator grant for TeraSHAPE (#773248); by the Deutsche Forschungsgemeinschaft (DFG) projects PACE (#403188360) and GOSPEL (#403187440) within the Priority Programme "Electronic-Photonic Integrated Systems for Ultrafast Signal Processing" (SPP 2111), by the DFG project HIPES (#383043731); by the project TeraSlice (#863322) under European Union's Horizon 2020 research and innovation programme, by the Alfried Krupp von Bohlen und Halbach-Stiftung; by the Karlsruhe School of Optics & Photonics (KSOP); by the Karlsruhe Nano Micro Facility (KNMF); by the European Regional Development Fund (ERDF), by the National Science Foundation (NSF) (DMR-1303080); and by the Air Force Office of Scientific Research (AFOSR) (FA9550-19-1-0069).

**Acknowledgment**. The authors would like to extend thanks to Jochen Schäfer for help with the VNA calibration, and to Keysight Technologies for borrowing us the 1 mm calibration kit.

See Supplementary Information 1 for more details

# Supplementary Information

This document provides supplementary information to "Hybrid Electro-Optic Modulator Combining Silicon Photonic Slot Waveguides with High-k Radio-Frequency Slotlines". Section I describes the fabrication of capacitively-coupled silicon-organic hybrid (CC-SOH) Mach-Zehnder modulators (MZM) as well as the optical and RF characterization of the underlying amorphous $BaTiO_3$ films. Section II gives more details on the measurement of electro-optic (EO) bandwidth of the CC-SOH MZM along with a derivation of analytical relations that connect the measured electrical $S$-parameters to the electro-optic transfer function of the device. Further, various factors that contribute to the bandwidth limitation of our first-generation CC-SOH MZM are discussed along with improvements to extend its bandwidth beyond 100 GHz.

## 1. Fabrication of CC-SOH MZM and characterization of $BaTiO_3$ thin films

The CC-SOH MZM used in our experiments is fabricated on a standard silicon-on-insulator (SOI) substrate, featuring a 220 nm-thick silicon (Si) device layer and a 2 μm-thick buried oxide ($SiO_2$) layer. The device is fabricated in a five-step lithographic process where the structures are defined by high-resolution electron-beam (e-beam) lithography. In the first lithography step, gold markers are fabricated on the SOI chip to ensure alignment accuracy of better than 50 nm between different exposures. In two subsequent lithographic steps, we fabricate shallow-etched grating couplers and fully-etched Si nanowire waveguides along with strip-to-slot converters. In the fourth lithographic step, ground-signal-ground (GSG) transmission lines made of 150-nm-thick gold are fabricated via a lift-off process using a poly-methyl methacrylate (PMMA) and an electron-beam-evaporated gold layer. In the last step, the high-k dielectric $BaTiO_3$ (BTO) is deposited and structured via a second lift-off process using PMMA as a mask. For the current device generation, amorphous BTO films with a thickness of 150 nm are deposited by room-temperature RF magnetron sputtering using a stoichiometric $BaTiO_3$ disk with 2-inch diameter as a target. The deposition is carried out at a base pressure of $5 \times 10^{-8}$ mbar, and the working pressure is set to $2.5 \times 10^{-3}$ mbar by controlling the flow rate (30 sccm) of Ar gas. A rather low RF power of 30 W is applied to the target to keep the temperature of SOI substrates below the glass transition temperature of the PMMA mask to enable a proper lift-off. Due to the low process temperature, the BTO thin film is known to be amorphous [1,2]. For the photonic structures, we measure an insertion loss of about 5 dB per grating coupler, a phase-shifter propagation loss of 6 dB/mm, and an additional 3 dB of loss from other passive silicon structures such as waveguides and MMI couplers. This results in a total fiber-to-fiber insertion loss of about 19 dB for a 1 mm-long CC-SOH MZM. The losses can be greatly reduced by optimized fabrication processes, which lead to smaller sidewall roughness of the slot waveguides, thereby bringing down the propagation losses to less than 0.2 dB/mm [3]. In combination with low-loss strip-to-slot converters [4] and MMI couplers [5], on-chip insertion losses of less than 1 dB may eventually be reached for fully optimized CC-SOH devices. Efficient fiber-chip coupling can, e.g., be accomplished by using 3D-printed micro-lenses [6] or photonic wire bonds [7,8].

### A. Complex refractive index of the BTO film in the near IR range

The optical properties of BTO thin films are determined from 150 nm-thick BTO layers deposited on $SiO_2$/Si substrates using the same process parameters that are used for fabricating the CC-SOH modulators as described in the previous section. By carrying out spectroscopic ellipsometric measurements, we determine a real part of the refractive index $n_{BTO} = 1.85$ along with a negligible imaginary part (extinction coefficient) $n_{i,BTO}$ at 1550 nm, see Fig. S1. The refractive index extracted from our measurements shows a good agreement with previously published values for amorphous $BaTiO_3$ layers deposited by room-temperature sputtering [9]. The slight differences in the indices is attributed to variations of the process parameters such as gas pressure in the deposition chamber, which results in composition variations of the deposited material.

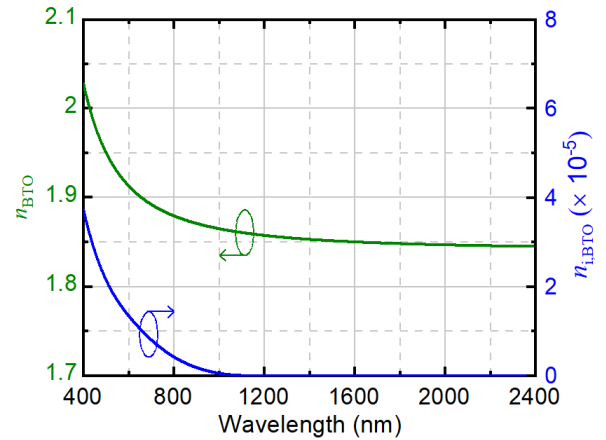

**Fig. S1:** Refractive index $n_{BTO}$ and extinction coefficient $n_{i,BTO}$ of 150 nm-thick amorphous BTO film measured using spectroscopic ellipsometry.

## B. Complex permittivity of the BTO film at RF frequencies

As mentioned in Section 2 of the main paper, the optical and RF properties of BTO are crucial for the performance of CC-SOH modulators. It is well known [10] that the material properties of BTO vary substantially depending on the deposition technique and its parameters. To determine the RF properties of our BTO films, we fabricate two sets of coplanar waveguides (CPW) with lengths $L$ of 1 mm, 2 mm, and 3 mm and a gap width of $w = 1\,\mu\text{m}$ between the ground and the signal traces, see Fig. S2. The first set of CPW only consists of the metal transmission line strips without any BTO, Fig. S2(b), whereas the second set contains a BTO thin film filling the gaps between the CPW strips, Fig. S2(c). For better comparability to the CC-SOH modulators described in the main paper, we used the same SOI wafers for these test structures and removed the silicon device layer prior to depositing the metal transmission lines and the BTO films. We measure the two-port $S$-parameters for all variations of the CPW using a vector network analyzer (VNA, Keysight PNA-X N5247) in the frequency range from 0.01 GHz to 110 GHz. The measured $S$-parameters are de-embedded using the transfer matrices of the fixtures that comprise on-chip RF contact pads and tapered transitions. From these measurements, we extract the respective complex propagation constant $\underline{\gamma}_{\text{BTO}}$ and $\underline{\gamma}_{\text{air}}$ as well as the complex line impedance $\underline{Z}_{\text{BTO}}$ and $\underline{Z}_{\text{air}}$ of the CPW with BTO-filled and air-filled gaps, respectively. We then determine the relative permittivity $\varepsilon_{\text{r,BTO}}$ of the amorphous BTO film using two different techniques – conformal mapping and an equivalent-circuit model.

### Conformal mapping technique

The conformal mapping technique (CMT) [11,12] is used to obtain closed-form expressions for calculating the characteristic impedance and the effective dielectric constant of a CPW structure. To use CMT, we assume [13] the metal strips of the CPW to be infinitely thin and perfectly conducting and to support a quasi-static transverse-electric (TEM) mode propagating along the line. We further assume [13] perfect magnetic walls at all dielectric boundaries and at the surface of the metal conductors, implying electric field lines are locally perpendicular to these boundaries. It is to be noted that CMT assumes dielectric layers with uniform thicknesses above and below the metal CPW, which is not the case in our structures having the BTO only in the gaps of the CPW, but not on the top of metals, see Fig. S2(c). However, since the electric field outside the BTO film is rather small, the associated error and the resulting overestimation of the dielectric constant $\varepsilon_{\text{r,BTO}}$ should be acceptable. The contribution of each of the various dielectric layers to the total effective permittivity of the CPW mode is then calculated using the so-called fill factors $q_i$ [13] that are given by

$$q_i = \frac{1}{2} \frac{K(k_i)}{K(k_i')} \frac{K(k_0')}{K(k_0)}, \quad i = \begin{cases} 1 & \text{for BTO} \\ 2 & \text{for SiO}_2 \\ 3 & \text{for Si} \end{cases} \quad \text{(S1)}$$

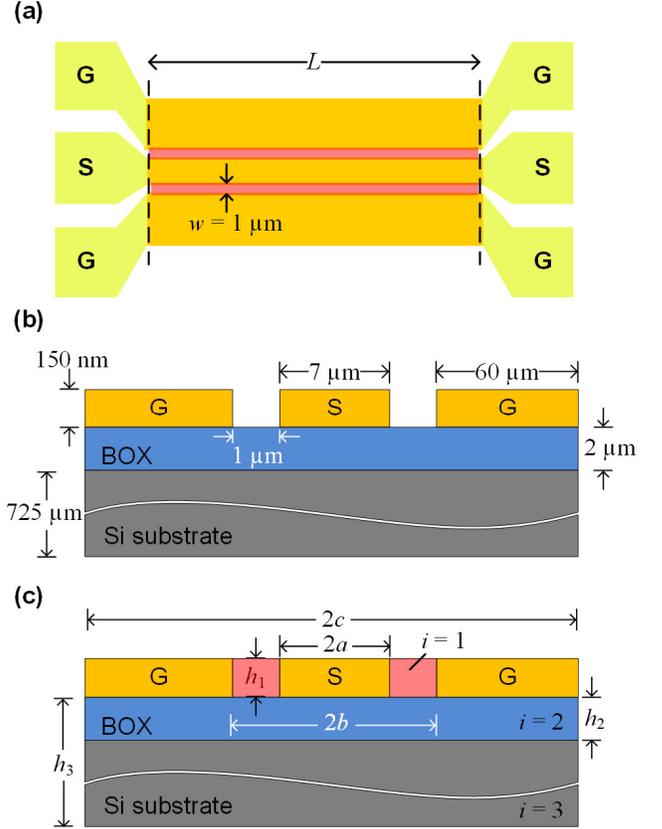

**Fig. S2:** CPW structures used for extracting the relative permittivity and the loss tangent of amorphous BTO films. **(a)** Top-view schematic of the CPW test transmission line with length $L$ and electrode spacing $w = 1\,\mu\text{m}$. **(b)** Cross-section of air-filled CPW test structure showing the geometric dimensions. **(c)** Cross-section of BTO-filled CPW showing the geometric parameters used in the conformal mapping analysis for extraction of relative permittivity of the BTO thin films.

In this relation, the function $K(x)$, $x = \{k_i, k_0, k_i', k_0'\}$, is the complete elliptic integral of the first kind (Section 17.3 in [14]) with

$$k_i = \frac{\sinh(\pi a/2h_i)}{\sinh(\pi b/2h_i)} \sqrt{\frac{1 - \sinh^2(\pi b/2h_i)/\sinh^2(\pi c/2h_i)}{1 - \sinh^2(\pi a/2h_i)/\sinh^2(\pi c/2h_i)}}, \quad \text{(S2)}$$

$$k_0 = \frac{a}{b}\sqrt{\frac{1 - b^2/c^2}{1 - a^2/c^2}}, \quad k_i' = \sqrt{1 - k_i^2}, \quad k_0' = \sqrt{1 - k_0^2}. \quad \text{(S3)}$$

Assuming relative permittivities for the oxide layer, $\varepsilon_{\text{r,SiO}_2} = 3.9$, and for the silicon layer, $\varepsilon_{\text{r,Si}} = 11.9$, the effective permittivity $\varepsilon_{\text{eff}}$ of the CPW mode in Fig. S2(c) is then given by [13,15]

$$\varepsilon_{\text{eff}} = 1 + q_1(\varepsilon_{\text{r,BTO}} - 1) + q_2(\varepsilon_{\text{r,SiO}_2} - \varepsilon_{\text{r,Si}}) + q_3(\varepsilon_{\text{r,Si}} - 1). \quad \text{(S4)}$$

From the frequency-dependent complex propagation constant $\underline{\gamma}_{\text{BTO,m}} = \alpha_{\text{BTO,m}} + j\beta_{\text{BTO,m}}$ extracted from the measurement of the

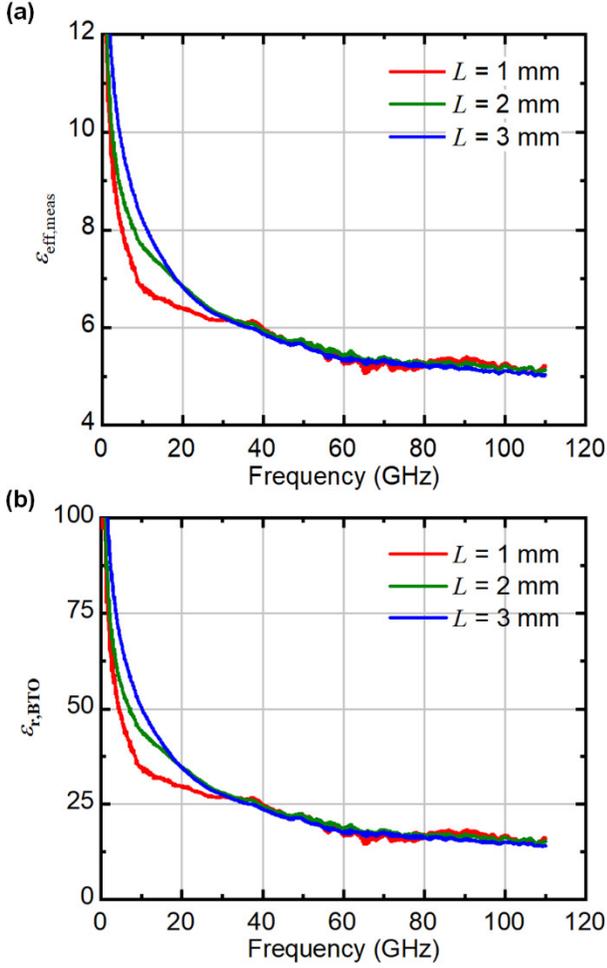

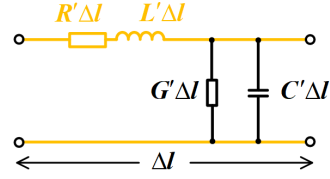

**Fig. S4:** Equivalent-circuit model of an infinitesimally short CPW segment of length $\Delta l$ used for estimating the relative permittivity $\varepsilon_{r,BTO}$ of the BTO layers along with the associated RF loss parameter $\tan \delta_{BTO}$.

**Fig. S3:** Measured frequency-dependent effective permittivity of the CPW and extracted relative permittivity of BTO. **(a)** Measured effective permittivity $\varepsilon_{eff,meas}(f)$ of the BTO-filled CPW for different lengths $L$. **(b)** Relative permittivity of BTO $\varepsilon_{r,BTO}(f)$ calculated from $\varepsilon_{eff,meas}(f)$ using conformal mapping, see Eqs. (S1) … (S5).

CPW S-parameters, the effective permittivity $\varepsilon_{eff,meas}$ of the mode is given by

$$\varepsilon_{eff,meas}(f) = \left( \frac{\beta_{BTO,m}}{2\pi f / c_0} \right)^2. \tag{S5}$$

In this relation, $c_0$ denotes the vacuum speed of light. The frequency-dependent relative permittivity of BTO $\varepsilon_{r,BTO}(f)$ can then be calculated by solving Eq. (S4) for $\varepsilon_{r,BTO}$ and by replacing $\varepsilon_{eff}$ by the measured values from Eq. (S5). The measured $\varepsilon_{eff,meas}(f)$ and the calculated $\varepsilon_{r,BTO}(f)$ obtained from transmission lines of different lengths $L$ are shown in Fig. S3(a) and (b), respectively. The results from the different transmission lines are in good agreement, confirming the validity of the approach.

**Equivalent-circuit model**

To confirm the results obtained from the CMT approach described in the previous section, we also model the air-filled and the BTO-filled CPWs, see Fig. S2(b) and (c), by the telegrapher's equations. Figure S4 shows an equivalent-circuit model of a infinitesimally short CPW section of length $\Delta l$. The frequency-independent elements $X' \in \{R', L', C', G'\}$ represent the resistance, the inductance, the capacitance and the conductance per unit length, respectively, and stand for either an air-filled $(X' = X'_{air})$ or an BTO-filled CPW $(X' = X'_{BTO})$. In terms of these quantities, the complex propagation constant and the complex line impedance of the CPW can be expressed as

$$\gamma = \alpha + j\beta = \sqrt{(R' + j\omega L')(G' + j\omega C')}, \tag{S6}$$

$$\underline{Z} = \sqrt{\frac{(R' + j\omega L')}{(G' + j\omega C')}}. \tag{S7}$$

To extract the various parameters $X'_{air}(X'_{BTO})$ for the air-filled (BTO-filled) CPW, we use a least-squares fit of Eqs. (S6) and (S7) to the complex propagation constant $\underline{\gamma}_{air}(\underline{\gamma}_{BTO})$ and to the complex impedance $\underline{Z}_{air}(\underline{Z}_{BTO})$ which were obtained from the measured complex S-parameters of the respective transmission lines. Note that this approach relies on the assumption of frequency-independent $X'_{air}(X'_{BTO})$ and hence does not allow to account for dispersive dielectrics with frequency-dependent permittivity $\varepsilon_{r,BTO}$. Except for the BTO filling, both the sets of transmission lines have geometrically identical cross-sections such that the partial capacitance $\Delta C'_{BTO}$ contribution of BTO thin film is given by

$$\Delta C'_{BTO} = C'_{BTO} - C'_{air}. \tag{S8}$$

From $\Delta C'_{BTO}$, the relative permittivity of BTO can be extracted in two ways — either by a parallel-plate capacitor approximation leading to $\varepsilon_{\mathrm{r,BTO(PP)}}$, or by the partial capacitance contribution of BTO calculated from CMT leading to $\varepsilon_{\mathrm{r,BTO(CMT)}}$,

$$\varepsilon_{\mathrm{r,BTO(PP)}} = \frac{C'_{\mathrm{BTO,P}}}{\varepsilon_0 \dfrac{h}{w}} \quad (S9)$$

$$\varepsilon_{\mathrm{r,BTO(CMT)}} = \frac{C'_{\mathrm{BTO,P}}}{2\varepsilon_0 \dfrac{K(k_1)}{K(k'_1)}} + 1. \quad (S10)$$

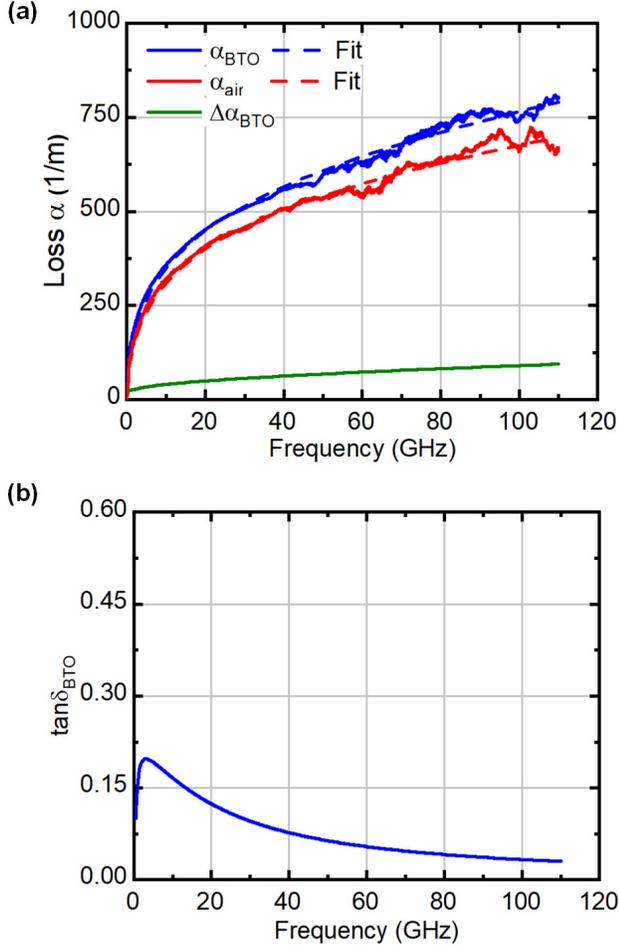

**Fig. S5:** Loss measurements of the air-filled and BTO-filled CPW and the extracted loss parameter $\tan\delta_{\mathrm{BTO}}$. **(a)** Measured (solid lines) and fitted (dashed lines) amplitude attenuation parameter $\alpha_{\mathrm{air}}(\alpha_{\mathrm{BTO}})$ of the air-filled (BTO-filled) CPW. The difference $\Delta\alpha_{\mathrm{BTO}}$ of the two quantities corresponds to the contribution of the lossy BTO film towards the overall propagation loss of the CPW. **(b)** Extracted loss tangent $\tan\delta_{\mathrm{BTO}}$ of the 150-nm-thick amorphous BTO film according to Eq. (S12).

With the parallel-plate capacitor approximation, we calculate a relative permittivity $\varepsilon_{\mathrm{r,BTO(PP)}} = 20$, while the CMT procedure leads to $\varepsilon_{\mathrm{r,BTO(CMT)}} = 18$. We attribute this slight deviation to the simplifying approximations that are associated with both techniques and to the fact that the fringing fields at the edge of the BTO region are not correctly considered. Still, the extracted permittivities are in good agreement with previously published values from [1,16–18] as well as with the CMT-calculated $\varepsilon_{\mathrm{r,BTO}}(f)$ shown in Fig S3(b).

To estimate the loss tangent of the BTO film, we compare the measured amplitude transmission factor $|\underline{S}_{21,\mathrm{air}}| = \exp(-\alpha_{\mathrm{air}}L)$ of an air-filled CPW with that of a BTO-filled, but otherwise identical CPW $|\underline{S}_{21,\mathrm{BTO}}| = \exp[-(\alpha_{\mathrm{BTO}})L]$. Figure. S5(a) depicts the measured (solid lines) amplitude loss coefficients $\alpha_{\mathrm{BTO}}$ and $\alpha_{\mathrm{air}}$ as a function of frequency along with fitted curves (dashed lines) according to Eqs. (S6) and (S7) The frequency-dependent contribution $\Delta\alpha_{\mathrm{BTO}}$ of the BTO layer towards the overall amplitude attenuation parameter $\alpha_{\mathrm{BTO}}$ of the BTO-filled line is then obtained by the difference of the fitted model functions,

$$\Delta\alpha_{\mathrm{BTO}} = \alpha_{\mathrm{BTO}} - \alpha_{\mathrm{air.}} \quad (S11)$$

The frequency-dependence of $\Delta\alpha_{\mathrm{BTO}}$ is shown as a green line in Fig. S5(a). The frequency-dependent loss tangent $\tan\delta_{\mathrm{BTO}}$ is then calculated by using Eq. (2.81) in Ref. [15] with parameters calculated or measured in the previous sections,

$$\tan\delta_{\mathrm{BTO}} = \frac{2c_0\Delta\alpha_{\mathrm{BTO}}}{\omega} \frac{\sqrt{\varepsilon_{\mathrm{eff,meas}}(f)}}{\varepsilon_{\mathrm{r,BTO(CMT)}}} \left( \frac{\varepsilon_{\mathrm{r,BTO(CMT)}} - 1}{\varepsilon_{\mathrm{eff,meas}}(f) - 1} \right). \quad (S12)$$

The extracted material loss parameter $\tan\delta_{\mathrm{BTO}}$ as shown in Fig. S5(b) agrees well with the estimated loss tangent values of amorphous BTO films as, e.g., published in [1,19].

## 2. EO figure-of-merit and bandwidth measurement of CC-SOH MZM

### A. Estimation of the EO figure-of-merit

As described in the Section 3 of the main paper, we measure a $U_\pi L$ product of 1.3 V mm for a CC-SOH MZM with 1 mm-long phase shifters. According to Eq. 7 in [20], the measured $U_\pi L$-product of the MZM is related to the EO figure-of-merit (FoM) $r_{33}$ of the organic EO material by

$$U_\pi L = \frac{w_s \lambda}{2 n_{\mathrm{EO}}^3 r_{33} \Gamma_s}. \quad (S13)$$

In this relation, $\lambda = 1550$ nm is the operating wavelength in vacuum, $w_s = 100$ nm is the slot width, $n_{\mathrm{EO}} = 1.75$ is the refractive index of the EO polymer YLD124, and $\Gamma_s$ denotes the field interaction factor that describes the interaction between the modulating RF field and the optical field with in the silicon slot waveguide, see Eq. (S6) in the supplementary information of [21]. Using an electromagnetic mode solver (CST MicrowaveStudio), we calculate $\Gamma_s = 0.32$ by assuming a relative permittivity $\varepsilon_{r,\text{BTO}} = 18$ for the BTO thin film. From this, we estimate an EO coefficient $r_{33} \approx 34\,\text{pm/V}$, which is still far below the EO coefficient of 390 pm/V that is possible for SOH devices when using more efficient EO materials such as JRD1, possibly in combination with slightly wider slots [21].

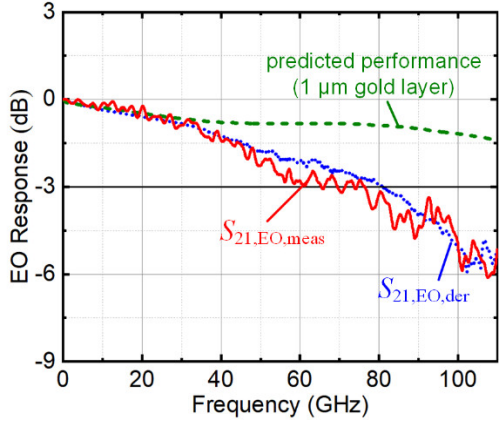

**Fig. S6:** EO response of a CC-SOH MZM having a phase shifter length $L = 1\,\text{mm}$. The red (solid) curve shows the directly measured EO response $S_{21,\text{EO,meas}}$, while the blue (dotted) curve represents the EO response $S_{21,\text{EO,der}}$ derived from the electrical $S$-parameters. The directly measured and the derived characteristics are in good agreement. The measured device is based metal transmission lines that are only 150 nm thick. The green (dashed) line represents the possible EO response of the CC-SOH MZM that can be achieved with an optimized device configuration having 1 µm-thick metal transmission lines that features smaller RF losses.

### B. Bandwidth of CC-SOH Mach-Zehnder modulator

The frequency response of the CC-SOH MZM is measured in the frequency range from 0.01 GHz to 110 GHz using a vector network analyzer (VNA, Keysight PNA-X N5247). A calibration kit for 1 mm-connectors is used to shift the measurement reference plane from the network analyzer ports to the end of the connected coaxial cables using a short-open-load-through (SOLT) calibration procedure. The RF signal from the VNA is coupled to the ground-signal-ground (GSG) pads of the CC-SOH MZM using a GSG probe (Cascade Infinity Probe I110-A-GSG-100). The other end of the MZM transmission line is terminated by a second probe connected to a $50\,\Omega$ impedance. We exploit the unbalanced arm lengths of the MZM to adjust the device to the quadrature operating point by tuning the wavelength of the feed laser. The intensity-modulated output of the CC-SOH MZM is detected by a calibrated high-speed photodiode (HHI, C05-W36), and the output signal is recorded by another port of the VNA. The photodiode has a 3 dB bandwidth of 78 GHz, followed by a smooth roll-off, which allows to perform measurements up 110 GHz, where the transfer function drops by 9 dB with respect to the low-frequency range. The EO response $S_{21,\text{EO,meas}}$ of the CC-SOH MZM can be de-embedded from the measured overall transmission $S_{21,\text{overall}}$ by taking into account the frequency response of the photodiode and of the RF probes, leading to the red curve in Fig. S6, which is equivalent to Fig. 2(d) of the main paper. For a 1 mm-long CC-SOH MZM, we measure a 3 dB EO bandwidth of 76 GHz.

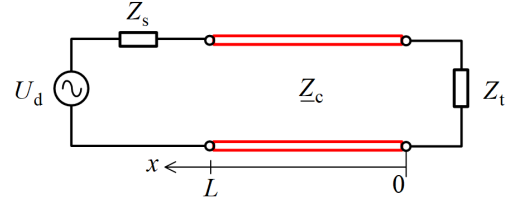

**Fig. S7:** Equivalent-circuit model for a traveling-wave modulator having a length $L$ and a complex line impedance $\underline{Z}_c$. The device is driven by an RF source with an internal impedance $Z_s$ and a voltage amplitude $U_d$. $Z_t$ denotes the termination impedance.

To confirm the directly measured EO dynamics, we also estimate the EO transfer function from the electrical scattering parameters. To this end, we measure the electrical $S$-parameters by connecting the second port of the VNA to the RF probe that was earlier used as a termination of the CPW. For estimating the EO response of the traveling-wave modulator, we use the equivalent circuit shown in Fig. S7. We assume a device of length $L$ and a complex RF impedance $\underline{Z}_c$, driven by an RF source of internal impedance $Z_s$ at RF modulation frequency $\omega_m / 2\pi$ and voltage amplitude $U_d$, and terminated by an impedance $Z_t$. The instantaneous voltage seen by the optical signal at a particular position on the line is then given by Eqs. 1 – 5 in [22],

$$U(x, \omega_m) = \frac{U_d}{2} \cdot (1 + \rho_1) \cdot e^{i\beta_o L} \cdot \frac{e^{i(\beta_e - \beta_o)x} + \rho_2 \cdot e^{-i(\beta_e + \beta_o)x}}{e^{i\beta_e L} + \rho_1 \cdot \rho_2 \cdot e^{-i\beta_e L}} \quad \text{(S14)}$$

$$\rho_1 = \frac{\underline{Z}_c - Z_s}{\underline{Z}_c + Z_s} \quad \text{(S15)}$$

$$\rho_2 = \frac{Z_t - \underline{Z}_c}{Z_t + \underline{Z}_c} \quad \text{(S16)}$$

$$\beta_o = \frac{\omega_m}{c_0} n_{g,\text{opt}} \quad \text{(S17)}$$

$$\beta_e = \frac{\omega_m}{c_0} n_{e,RF} - i \cdot \alpha_m \qquad (S18)$$

In these relations, $\alpha_m$ is the amplitude attenuation constant of the RF signal, $n_{g,opt}$ denotes the optical group refractive index, $n_{e,RF}$ is the effective index of the RF signal, and $c_0$ is the speed of light in vacuum. The phase shift experienced by an optical wave traveling through the modulator is directly proportional to the voltage experienced by it along the length of the phase shifter. To simplify the analysis, it is convenient to work with the average voltage along the length $L$ of the modulator, see Eq. 17 in Ref. [23],

$$U_{avg}(\omega_m) = \frac{1}{L}\int_0^L U(x,\omega_m)\,dx$$
$$= \frac{U_d \cdot (1+\rho_1) \cdot e^{i\beta_o L}}{2(e^{i\beta_e L} + \rho_1 \cdot \rho_2 \cdot e^{i\beta_e L})} \cdot (U_+ + \rho_2 \cdot U_-), \qquad (S19)$$

where

$$U_\pm = e^{\pm i\phi_\pm} \frac{\sin\phi_\pm}{\phi_\pm}, \quad \phi_\pm = \frac{(\beta_e \mp \beta_o)L}{2}. \qquad (S20)$$

Note that Eq. (S19) accounts for the impact of RF propagation loss, impedance mismatch of the modulator with respect to source and termination impedance, as well as velocity mismatch of the RF and the optical waves of the traveling-wave modulator.

The frequency response $m(\omega_m)$ of the modulator is finally given by the frequency-dependent phase amplitude that is normalized to its value at zero angular frequency. Taking into account that the voltage effective at the slot waveguide is reduced with respect to the average voltage $U_{avg}(\omega_m)$ on the transmission line, the EO frequency response can be written as

$$m(\omega_m) = \left| H(\omega_m) \cdot \frac{U_{avg}(\omega_m)}{U_{avg}(0)} \right|, \qquad (S21)$$

where the transfer function $H(\omega_m)$ represents the voltage divider formed by series combination of the coupling capacitance $C_c$ of the BTO and the slot capacitance $C_s$, see Fig. 1(c) of the main paper,

$$H(\omega_m) = \frac{C_c(\omega_m)}{C_c(\omega_m) + C_s}. \qquad (S22)$$

In a parallel-plate approximation, the capacitances $C_c$ and $C_s$ can be estimated from the geometrical parameters of the MZM and the relative permittivities of the various materials. For the BTO thin film, we use the frequency-dependent relative permittivity $\varepsilon_{r,BTO}(\omega_m)$ derived in the Section 1B, see Fig. S3(b), and use a frequency-independent dielectric constant of $\varepsilon_{r,EO} = 5.68$ for the EO polymer that fills the silicon slot.

For calculating the EO frequency response, we extract the complex characteristic impedance $\underline{Z}_c$, the RF amplitude attenuation coefficient $\alpha_m$, and the RF effective index $n_{e,RF}$ of the CC-SOH MZM from the electrical S-parameters of the device. The optical group refractive index $n_{g,opt} = 2.8$ needed in Eq. (S17) is numerically calculated using a commercially available optical mode solver (CST MicrowaveStudio). The transfer function of the RF fixtures, i.e., the on-chip metal contact pads and the tapered transitions to the metal strips of the transmission line, are de-embedded by using S-parameter measurements of CC-SOH MZM of different modulator lengths. We finally obtain the dB-values of the EO response $S_{21,EO,der}$ derived from the electrical scattering parameters of the CC-SOH MZM as $20\log_{10}(m(\omega_m))$. For a device with a phase shifter length of $L = 1\,\text{mm}$, the results are indicated by a blue dotted line in Fig. S6. The EO response $S_{21,EO,der}$ derived from the electrical scattering parameters is in good agreement with its directly measured counterpart $S_{21,EO,meas}$, thus confirming the validity of the approach.

Furthermore, we investigate the limiting factors for the bandwidth of the CC-SOH MZM. From the numerically calculated value of $n_{g,opt} = 2.8$ and the mean RF effective index $\bar{n}_{e,RF} = 2.2$ averaged over the frequency range 0.01 GHz … 110 GHz, the velocity-mismatch-limited 3dB-bandwidth of a CC-SOH device is given by Eq. 2 in Ref. [24],

$$f_{vm} = 1.4 \frac{c_0}{\left(\pi \left| n_{g,opt} - \bar{n}_{e,RF} L \right| \right)}. \qquad (S23)$$

For our 1 mm-long CC-SOH devices, we estimate a velocity-mismatch-limited 3 dB frequency $f_{vm} = 220\,\text{GHz}$. It is thus clear that the bandwidth limitation of 76 GHz observed in our measurements cannot be caused by velocity mismatch, but arises from the non-optimum electrical design of our first-generation device, especially from the high RF propagation loss. Specifically, the metal transmission line is fabricated using a lift-off process with a thin photoresist, which limited the thickness of the gold layer to a rather small value of approximately 150 nm. This leads to significant RF propagation loss, which increases strongly with frequency. This RF propagation loss in combination with a frequency-dependent line impedance remains as the most important reason for the strong decay in the frequency response of our first-generation CC-SOH modulators.

The RF propagation loss has two main contributions – ohmic loss and dielectric loss. As discussed in the Section 1B above, the dielectric loss of the BTO thin film is negligible, and the RF propagation loss is dominated by the ohmic losses in the 150 nm-thick gold transmission lines. Specifically, from our characterization of the test structures described in the Section 1B, we find a BTO-related dielectric loss of only 0.58 dB/mm at 50 GHz, while the overall propagation loss of 4.5 dB/mm is significantly higher. For the modulator structures, we extract the frequency-dependent propagation loss from the electrical S-parameter measurements and compare them with the results of numerical simulations (CST Microwave Studio). We find very good agreement when assuming a conductivity of $\sigma_{Au} = 1.65 \times 10^7 \, \mathrm{S/m}$ for the 150-nm-thick gold strips, see red and blue curves in Fig. S8. Note that the conductivity value assumed for gold thin films is smaller than the conductivity of $4.5 \times 10^7 \, \mathrm{S/m}$ for bulk gold – a common phenomenon observed in thin gold layers depending on the process conditions [25]. We then repeat the simulation with an increased thickness of the gold layer of 1 μm, while still assuming the conductivity of $\sigma_{Au} = 1.65 \times 10^7 \, \mathrm{S/m}$ for the gold thin film. The increased gold thickness greatly reduces the RF propagation loss, see green curve in Fig. S8, and leads to an attenuation of approximately 1 dB/mm at 50 GHz. To study the overall performance of the CC SOH MZM with the 1 μm-thick gold layer, we use the simulated (CST Microwave Studio) frequency-dependent RF propagation loss along with the frequency-dependent line impedance to re-evaluate the electro-optic (EO) response according to the model described in Section 2B. This leads to a 3 dB EO bandwidth significantly larger than 100 GHz, see green curve in Fig. S6 and Fig. S2(d) of the main paper.

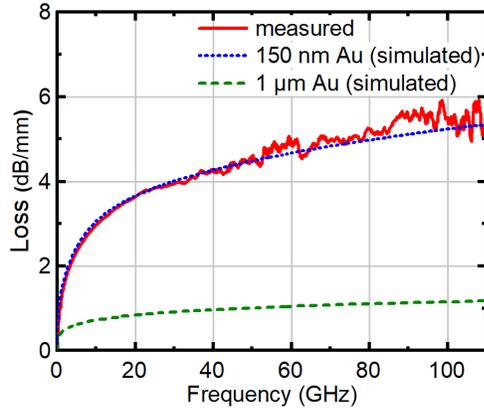

**Fig. S8:** Conductor loss of the measured (red solid curve) CC-SOH MZM and that of the simulated device (blue dotted curve) with 150 nm Au (gold) transmission line for a conductivity $\sigma_{Au} = 1.65 \times 10^7 \, \mathrm{S/m}$. The green dashed curve shows the reduction of the loss when using a 1 μm-thick gold transmission line.